\documentstyle[twoside,fleqn,npb,epsfig]{article}
\newcommand{\AmS}{{\protect\the\textfont2
  A\kern-.1667em\lower.5ex\hbox{M}\kern-.125emS}}

\title{Fragmentation in DIS } 

\author{L. Zawiejski  \\
for the ZEUS Collaboration 
\address{Institute of Nuclear Physics, \\
        Kawiory 26a, PL-30-055 Cracow, Poland}}

\begin{document}

\begin{abstract}

The fragmentation properties in deep-inelastic scattering (DIS) at HERA are investigated 
in order to test perturbative QCD and quark fragmentation universality.
Selected results for the inclusive single particle distributions and the angular 
two-particle correlations in the current region of the Breit frame are presented and compared
to results from  $e^{+}e^{-}$ collisions. The extension of the DIS fragmentation studies
to the target region of the Breit frame are also reported.
\end{abstract}

\maketitle
\section{INTRODUCTION}
Many aspects of fragmentation phenomenon in DIS have been extensively studied through 
the last years by ZEUS in terms of the single particle scaled momentum distributions 
and the energy evolution of their higher order moments, fragmentation functions~\cite{a1,a2},
the mean charged multiplicity energy dependence~\cite{a3} or transverse momentum  
spectra~\cite{a4}. These results were obtained in the current hemisphere 
of the Breit frame~\cite{a5}. 
Recently the measurements were extended also to
particle correlations~\cite{a6} and in some limited range to the target fragmentation 
region. Such DIS studies allowed to test universality of perturbative QCD together with LPHD,
universality of the quark fragmentation by comparison with  $e^{+}e^{-}$ data and to learn about
the dynamics related with proton fragmentation hemisphere. In addition, obtained results 
uncover the current and target region relationship which can be quantitatively estimated.   
\section{SINGLE PARTICLE SPECTRA}
\begin{figure}[htb]
\vspace{9pt}
\hspace*{-1.2cm}
\epsfig{file=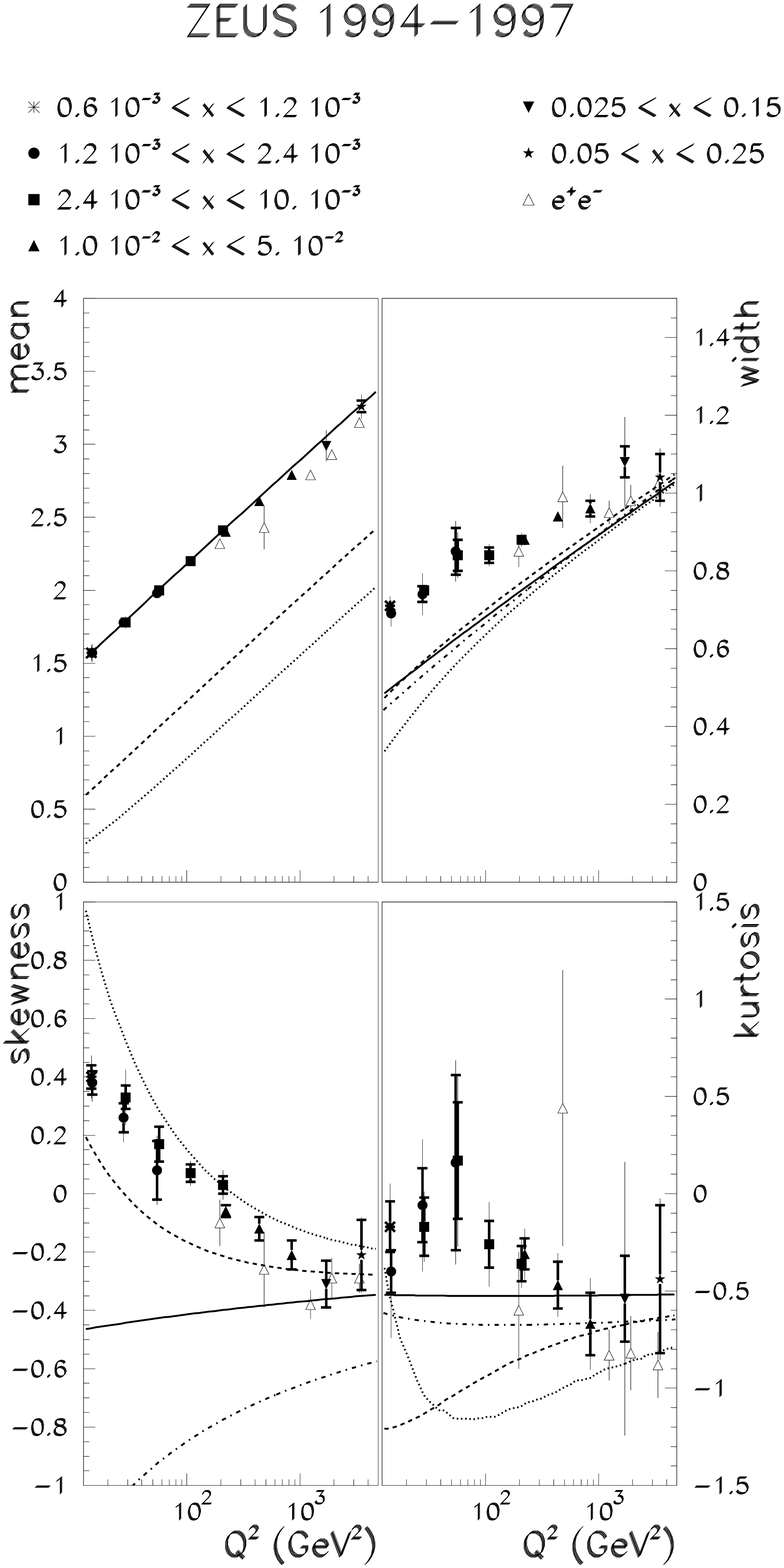,height=9.5cm,width=9.5cm}
\vspace*{-0.9cm}
\caption{The moments of $\xi$ distribution for DIS and $e^+e^-$ data compared to MLLA 
predictions~\protect\cite{a8} (the full line is $Q_0=\Lambda$, the dashed $Q_0=2\Lambda$,
the dotted $Q_0=3\Lambda$).
The dash-dotted line is prediction coming from ref~\protect\cite{a9}.}
\label{fig:p1}
\end{figure}

The scaled momentum distributions in terms of variable $ \xi=\ln(1/x_{p})$ 
($x_{p}=2p^{Breit}/Q$) in different $(x,Q^{2})$ bins
are approximatelly Gaussian distributions of $ \xi$
with  rising maximum and shifted peak position to higher $\xi$
with rising $Q^{2}$~\cite{a2}. Such behaviour is generally  consistent with
MLLA calculacions in which color coherence phenomenon~\cite{a7} was included. 
The available recently statistics for DIS events make it possible to perform more detailed
comparison of data with pQCD predicted energy evolution of higher moments of $\xi$ 
distributions.
These moments can be obtained from fits to distorted Gaussian
as supplied by MLLA calculations:
\begin{eqnarray}
{D(\xi)} \propto  
exp(\frac{1}{8}k-\frac{1}{2}s\delta-\frac{1}{4}(2+k)\delta^{2}+\nonumber\\
\frac{1}{6}s\delta^{3}+
\frac{1}{4}k\delta^{4})  
\end{eqnarray}
$\delta = (\ln(1/x_p)-l)/\sigma$ and $l$, $\sigma$, $s$ and  $\kappa$ are the 
mean value, the dispersion, the skewness and  the kurtosis of the $\xi$ distribution. 
Fig.~\ref{fig:p1} shows $Q^{2}$  evolutions of these moments for the DIS, $e^{+}e^{-}$ data
and the MLLA predictions~\cite{a8,a9}. The MLLA predictions together with LPHD do not
describe the data. The reasonable agreement DIS and  $e^{+}e^{-}$ results is consistent
with the universality of quark fragmentation. 
\vspace*{-0.2cm}
\section{ANGULAR CORRELATIONS}
\begin{figure}[htb]
\hspace*{-0.2cm}
\epsfig{file=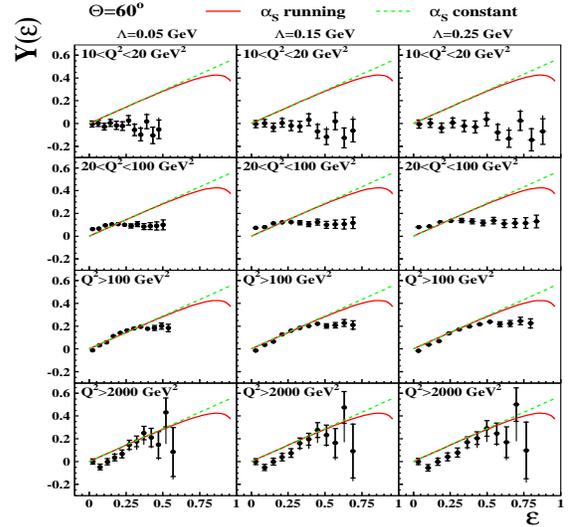,height=7.5cm,width=7.5cm}
\vspace*{-1.2cm}
\caption{The angular correlations and DLA predictions~\protect\cite{a10}.}
\label{fig:p2}
\end{figure}

The correlations between all pairs of charged particles in the
current region of Breit frame in a cone of half-opening angle $\Theta$ were calculated
according to following formulae~\cite{a10}:
\begin{eqnarray}
Y(\epsilon) = \frac {ln(\rho(\epsilon)/\rho_{mix}(\epsilon))}
                                         {\sqrt{ln(Psin\Theta/\Lambda)} }   
\end{eqnarray}
Such correlations were calculated and extensively studies in~\cite{a10}
where the scaling variable $\epsilon$ was introduced:
\begin{eqnarray}
\epsilon = \frac{ln(\Theta/\theta_{12})}{ln(P\sin\Theta/\Lambda)}
\end{eqnarray} 
where $P=Q/2$, $\Lambda$ is the QCD energy scale and $\theta_{12}$ is the relative 
angle between two particles in each found pair.
The $\rho(\epsilon)$ is inclusive two-particle density:
\begin{eqnarray}
\rho (\epsilon) =  \frac{1}{N_{ev}}\frac{dn_{pair}}{d\epsilon} 
\end{eqnarray}
and $\rho_{mix}(\epsilon)$ is calculated for particles from different events to remove
any dynamical correlations for particles in events.
Fig.~\ref{fig:p2} shows these results together with DLA predictions~\cite{a10}.
The DLA description fails at low $Q^2$ but improves with increasing $Q^2$.
\vspace*{-0.8cm} 
\vspace*{-1.cm}
\section{CURRENT-TARGET REGION}
\begin{figure}[htb]
\hspace*{-0.5cm}
\epsfig{file=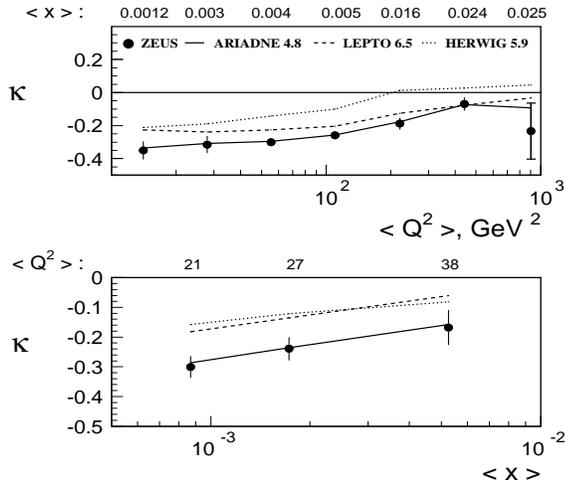,height=8.5cm,width=7.8cm}
\vspace*{-1.8cm}
\caption{The evolution of the correlation coefficient $\kappa$ with $Q^2$ and $x$.}
\label{fig:p3}
\end{figure}

In~\cite{a2}, the comparison of the scaled momentum, the transverse momentum 
and multiplicity spectra between the current and target region 
of the Breit frame were presented. The significant
differencies were found, these suggest 
the different dynamical mechanism in their origin.
According to the recent results~\cite{a11} the current and the target hemisphere  
can not be treated separately when first order QCD processes: boson gluon fusion
(BGF) and QCD Compton (QCDC) appear.
In this case the calculations lead to the current-target anticorrelations.
The experimental results confirm such predictions. The strenght of this effect can be 
supplied by the calculation of the correlation coefficient defined as follows:
\begin{eqnarray}
\kappa = \frac{< n_cn_t > - < n_c > < n_t>}
                      {\sigma_c\sigma_t} 
\end{eqnarray}
$n_c$ ($n_t$) is the number of particles in the current (target) region and
$\sigma_{c}$ ($\sigma_{t}$) is the standard deviation of the multiplicity distributions
in current (target) region. The value of $\kappa$ is limited: -1 $\leq \kappa \leq$ 1 
and in absence of the correlations $\kappa$ = 0.
Fig.~\ref{fig:p3} shows the dependence of the correlation coefficient $\kappa$ as
a function of $x$ and $Q^2$. Anti-correlations are getting stronger with descreasing $x$ 
and become weaker with increasing $Q^2$.
\vspace*{-1.5cm}
\vspace*{-1.cm}
\section{QUARK FRAGMENTATION UNIVERSALITY}
\begin{figure}[htb]
\hspace*{-0.2cm}
\epsfig{file=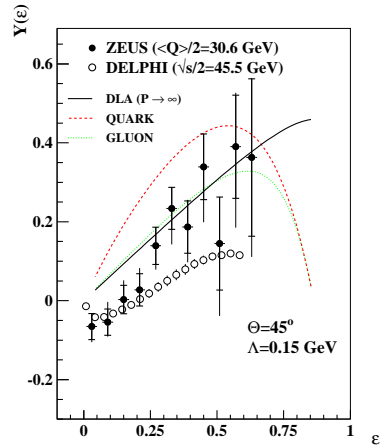,height=6.5cm}
\vspace*{-1.cm}
\caption{The angular correlations measured in DIS and $e^{+}e^{-}$ data~\protect\cite{a12}.
In addition the asymptotic DLA predictions~\protect\cite{a10} and those for final
energy of the initial quark or gluon.}
\label{fig:p4}
\end{figure}

The quark fragmentation universality was checked previously for single particle
spectra~\cite{a2} (fragmentation function, mean multiplicity) by comparison 
of DIS and  $e^{+}e^{-}$ data. The results support the universality of the quark
fragmentation. 
In Fig.~\ref{fig:p4} such comparison is presented for the angular correlations
$Y(\epsilon)$.
The disagreement for correlations between the DIS and  $e^{+}e^{-}$ data is observed.
The large part of these differencies can be explained by influence
of the first order QCD processes (BGF and QCDC) as the different jet
configurations in the current and target region would change
the number of migrated particles into each hemispheres. \\
\hspace*{0.1cm}
{\bf Acknowledgements:}  I would like to thank the DESY Directorate for the support. \\
Thanks also to my colleagues S. Chekanov, M. Przybycien and J. Chwastowski for discussions 
and critical reading this manuscipt.


\begin{thebibliography}{9}
\bibitem{a1} ZEUS Collaboration, J. Breitweg et al., Phys. Lett. B414 (1997) 428.
\bibitem{a2} ZEUS Collaboration, J. Breitweg et al., DESY 99-041 (submitted to Eur. Phys. J.C).
\bibitem{a3} ZEUS Collaboration, M. Derrick et al., Z. Phys. C67 (1995) 93.
\bibitem{a4} ZEUS Collaboration, M. Derrick et al., Z. Phys. C70 (1996) 1. 
\bibitem{a5} R.P. Feynman, "Photon-Hadron Interactions", Benjamin, New York, 1972. 
\bibitem{a6} ZEUS Collaboration, J. Breitweg et al., DESY 99-063 (submitted to Eur. Phys. J.C).
\bibitem{a7} Ya.A. Azimov, Yu.L. Dokshitzer, V.A. Khoze and S.I. Troyan, Z.Phys. C31 (1986) 213.
\bibitem{a8} Yu.L. Dokshitzer, V.A. Khoze and S.I. Troyan, Int. J. Mod. Phys. A7 (1992) 1875.
\bibitem{a9} C.P. Fong and B.R. Weber, Phys. Lett. B229 (1989) 289; \\
C.P. Fong and B.R. Weber, Nucl. Phys. B355 (1991) 54.
\bibitem{a10} W. Ochs and J. Wosiek, Z.Phys. C68 (1995) 269; private communications.
\bibitem{a11} S.V. Chekanov, J.Phys. G25 (1999) 59.
\bibitem{a12} DELPHI Collaboration, P. Abreu et al., Phys. Lett. B440 (1998) 203.
\end{thebibliography}
\end{document}